\documentclass[pdflatex,sn-mathphys-num]{sn-jnl}

\usepackage{graphicx}%
\usepackage{multirow}%
\usepackage{amsmath,amssymb,amsfonts}%
\usepackage{amsthm}%
\usepackage{mathrsfs}%
\usepackage[title]{appendix}%
\usepackage{xcolor}%
\usepackage{textcomp}%
\usepackage{manyfoot}%
\usepackage{booktabs}%
\usepackage{algorithm}%
\usepackage{algorithmicx}%
\usepackage{algpseudocode}%
\usepackage{listings}%
\usepackage{rotating} 
\usepackage{adjustbox}
\usepackage{pifont}

\theoremstyle{thmstyleone}%
\theoremstyle{thmstyletwo}%

\theoremstyle{thmstylethree}%

\begin{document}

\title[Article Title]{Agreement–Disagreement Guided Knowledge Transfer for
Cross-Scene Hyperspectral Imaging}


\author[1]{\fnm{Lu} \sur{Huo}}\email{lu.huo@uts.edu.au}


\author[1]{\fnm{Haimin} \sur{Zhang}}\email{haimin.zhang@uts.edu.au}
\author*[1]{\fnm{Min} \sur{Xu}}\email{min.xu@uts.edu.au}

\affil*[1]{\orgdiv{School of Electrical and Data Engineering}, \orgname{University of Technology Sydney}, \orgaddress{\street{15 Broadway}, \city{Sydney}, \postcode{2007}, \state{NSW}, \country{Australia}}}




\abstract{Knowledge transfer plays a crucial role in cross-scene hyperspectral imaging (HSI). However, existing studies often overlook the challenges of gradient conflicts and dominant gradients that arise during the optimization of shared parameters. Moreover, many current approaches fail to simultaneously capture both agreement and disagreement information, relying only on a limited shared subset of target features and consequently missing the rich, diverse patterns present in the target scene. To address these issues, we propose an Agreement–Disagreement Guided Knowledge Transfer (ADGKT) framework that integrates both mechanisms to enhance cross-scene transfer. The agreement component includes GradVac, which aligns gradient directions to mitigate conflicts between source and target domains, and LogitNorm, which regulates logit magnitudes to prevent domination by a single gradient source. The disagreement component consists of a Disagreement Restriction (DiR) and an ensemble strategy, which capture diverse predictive target features and mitigate the loss of critical target information. Extensive experiments demonstrate the effectiveness and superiority of the proposed method in achieving robust and balanced knowledge transfer across heterogeneous HSI scenes.}


\maketitle
\section{Introduction} \label{sec1}

Hyperspectral imaging (HSI) processing has garnered attention in the field of remote sensing since HSI provide a wealth of information by capturing reflectance data across hundreds of contiguous spectral bands \cite{ghamisi2017advances,Landgrebe2002HyperspectralID,flarence2025hyperspectral, dabas2025construction}. However, collecting labeled hyperspectral data is time-consuming and expensive, making it impractical to gather extensive labeled datasets for every new scene encountered \cite{ye2017dictionary}. To address this challenge, knowledge transfer enables models trained on source scene with abundant labeled data to be adapted to a target scene with limited labeled data \cite{day2017survey}.


Existing cross-scene knowledge transfer methods in HSI mainly focus on two scenarios: homogeneous, which addresses spectral shifts within the same scene \cite{deng2018active,ye2017dictionary,yu2021unsupervised}, and heterogeneous, which involve matching shared categories before aligning spectral distributions across different scenes \cite{peng2022domain,ning2023contrastive,ye2024adaptive,wang2024dual}. Despite progress in both homogeneous and heterogeneous knowledge transfer scenarios, two critical challenges remain. First, current methods often under-explore the impact of optimization issues in shared parameters, particularly when there are very few training samples for the target task. This oversight can degrade performance and lead to harmful transfer effects in the target scene \cite{zhang2022survey, rosenstein2005transfer,senushkin2023independent}. 
Second, current approaches overlook the importance of incorporating both agreement and disagreement in the transfer process. Although the target scene may contain a wealth of complex and informative features, the predictive outcomes often rely solely on a limited, shared subset of target features. This reliance on restricted information prevents the model from fully capturing the rich and diverse patterns present in the target scene \cite{pagliardini2023agree}.


In this work, we consider cross-scene knowledge transfer from two perspectives: agreement and disagreement. To achieve agreement, the shared parameters for knowledge transfer may encounter conflicting and dominating gradients \cite{senushkin2023independent,yu2020gradient}. The gradients derived from the source and target scenes can oppose each other, leading to conflicting gradient directions when the model attempts to optimize for both scenes simultaneously. In addition, source task may have a larger dataset compared to target task, leading to larger gradients that dominate the learning process, potentially resulting in underfitting for the target task. 
In cases of disagreement, when the model focuses on shared features between the source and target domains—which represent only a small subset of the predictive features for the target—it limits the diversity of the results. This approach risks losing important information that could be critical for the target task, especially if the data in the target scene differs significantly from the source scene.

To solve these challenge, we propose a method called agreement-disagreement guided knowledge transfer (ADGKT), which consists of two main components: agreement and disagreement mechanisms. The agreement part includes GradVac \cite{wang2020gradient} and LogitNorm \cite{wei2022mitigating}; GradVac adjusts the direction of gradients during training to align them more closely, reducing conflicts between source and target gradients, while LogitNorm prevents dominating gradients by controlling the magnitude of the logits (pre-softmax outputs), equalizing the contributions from each task and reducing the impact of dominating gradients when updating shared parameters. The disagreement component comprises a disagreement restriction and an ensemble mechanism; the disagreement restriction (DiR) promotes learning diverse representations of instances in the target domain, and the ensemble component captures a diverse set of predictive target features, mitigating the risk of losing important target information. Therefore, ADGKT enhances the model's ability to generalize effectively to the target scene.


Our contributions are summarized as follows:
\begin{itemize}
    \item We propose agreement mechanisms including GradVac and LogitNorm components to adjust the gradients between the source and target scenes, mitigating gradient conflicts and dominating gradients during optimization of shared parameters.
    \item We introduce disagreement mechanisms including disagreement restriction and an ensemble components to capture rich and diverse representations for target samples, thereby alleviating a risk of losing critical target information. 
    \item  By jointly addressing these
two aspects, our method captures comprehensive target scene information and achieves improved generalization.
\end{itemize}


\section{Method} \label{sec2}
\begin{figure}
    \centering
    \includegraphics[width=\linewidth]{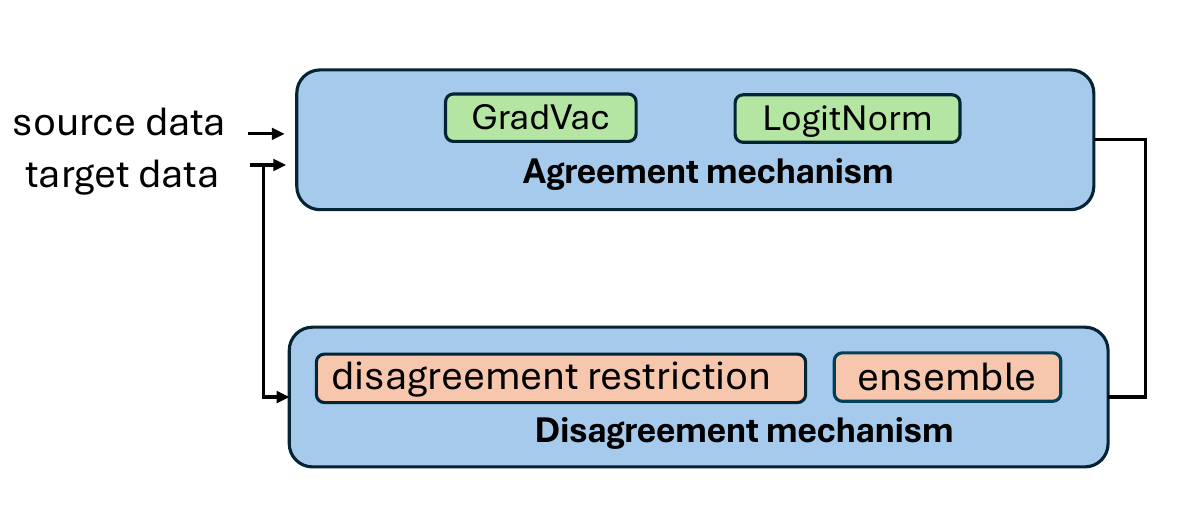}
    \caption{Our method contains two mechanisms including agreement and disagreement during the knowledge transfer from source scene to target scene. The agreement mechanism contain GradVac to alleviate gradient conflict and LogitNorm to mitigate the dominant gradient. Additionally, the disagreement mechanism comprise disagreement restriction to promote diversity and ensemble to capture agreement and disagreement aspects of target features.}
    \label{overview}
\end{figure}


\subsection{Problem Setting}
In cross-scene HSI knowledge transfer scenario, we define the source scene as $\mathcal{D}_s = \{(x^s_{i}, y^s_{i})\}_{i=1}^{N_s}$, where $x^s_{i} \in \mathbb{R}^{H_s \times W_s \times B_s}$ is a source scene input with \( H_s \) height, \( W_s \) width, and \( B_s \) spectral bands. The corresponding labels are denoted by \( y^s_{i} \in \mathcal{Y}_s \), where \( \mathcal{Y}_s \) is the label space for the source scene.
Similarly, let \( \mathcal{D}_t = \{(x^t_{j}, y^t_{j})\}_{j=1}^{N_t} \) represent the target scene, where \( x^t_{j} \in \mathbb{R}^{H_t \times W_t \times B_t} \) represents the target scene input. 
The target labels are represented by \( y^t_{j} \in \mathcal{Y}_t \), where \( \mathcal{Y}_t \) is the label space for the target scene. In addition, these two scenes have gradients $g_s$ and $g_t$ corresponding to their respective loss functions $\mathcal{L}_s$ and  $\mathcal{L}_t$.

\subsection{Technical Challenges}



One of the central issue in knowledge transfer between cross-scene HSI is achieving agreement in the optimization of shared parameters, which often suffers from gradient conflicts and dominating gradients.
During the model training, the optimization process often encounters gradient conflicts, where the gradients derived from source and target data pull the model’s parameters in different directions, hindering the learning process. 
Without effectively resolving these gradient conflicts, the model can converge to suboptimal solutions, and the transferred knowledge fail to generalize well to the target scene. 
In addition, the source task with abundance of training examples can dominate the update of shared network layers because it produces larger magnitude of gradients and gets more updates.
The model prioritize optimizing source task over the target task. The target task with fewer samples struggle to learn equally magnitude of logits, leading to potential underfitting.

Another issue during knowledge transfer is the failure to capture the rich and diverse information present in the target scene. The disagreement information for target scene is often neglected, especially when the source and target scenes exhibit different data characteristics.
During the agreement part, it is possible that neglect the target critical information since this information is less important in the source scene. Because of that, the model can overlook the target critical information, which is valuable for the target task. As a result, the final predictive feature for target scene might not transfer well from source to target scene. 
How to build effective transfer from source to target scene and make sure the integrity of target information remains a key challenge.



\subsection{Agreement-Disagreement Guided Knowledge Transfer (ADGKT)}

\begin{figure}[tb]
    \centering
    \includegraphics[width=\linewidth]{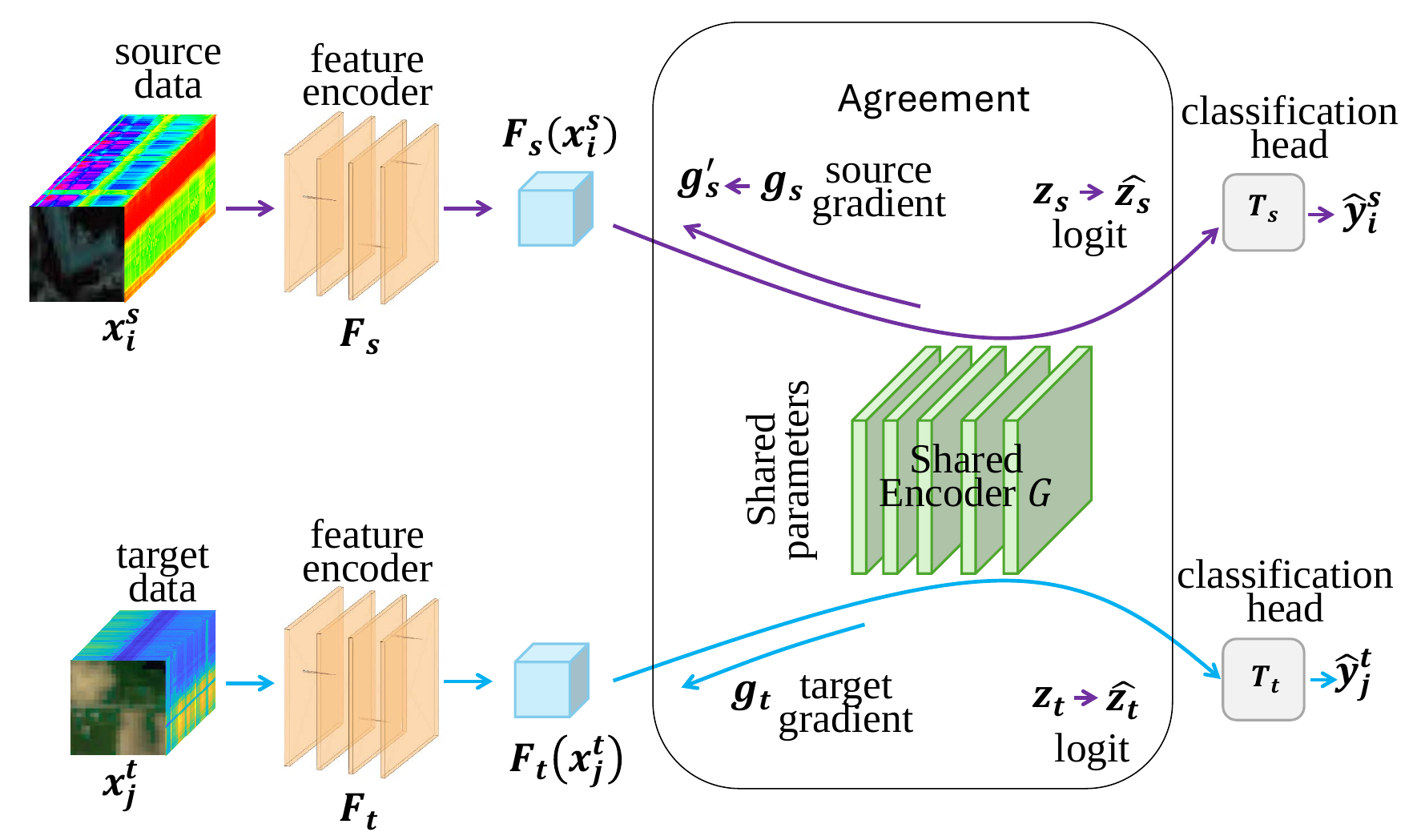}
    \caption{The agreement mechanism is proposed in our model. To alleviate gradient conflict and dominant gradient issues in shared encoder $G$, we employ GradVac to adapts the gradients $g_s$ to $g_s'$, thereby reducing gradient conflict. In addition, we utilize LogitNorm to get updated logits $\hat{z_s}$ and $\hat{z_t}$ , controlling their magnitude to prevent potential underfitting.    }
    \label{agreement}
\end{figure}

\begin{algorithm}[h]
    \caption{The pseudocode for the core of ADGKT framework}
    \label{alg:4}
    \begin{algorithmic}[1]
    \State{\textbf{Initialization:} Set $t=0$, the input data $x_{i}^s$ from HSI source scene and $x_{j}^t$ from HSI target scene. Specifically, $g_s$ for source scene and $g_t$ for target scene are the gradients to their respective loss functions $\mathcal{L}_s$ and  $\mathcal{L}_t$. $z$ is the logit before the softmax. 
    }
        \For{$t=0,1,2,\cdots,$}
            \State {\textbf{Agreement:}  
             \If{$\phi < \alpha$}
             \State $g_s' = g_s + \eta g_t$  \qquad
            //alleviating gradient conflict 
            \EndIf
            \If{$\Phi(g_s, g_t) \neq 1$}
            \State $\hat{{z}} = \frac{{z}}{\tau\|{z}\|}, \quad
            \mathcal{L}(\hat{z}_i, y) = - \log \left( \frac{\exp(\hat{z}_i)}{\sum_{j=1}^{C} \exp(\hat{z}_j)} \right)$
            
            \qquad//mitigating the dominating gradients
            \EndIf};
            \State {\textbf{Disagreement:}  

            $E_{\text{DiR}} = \mathbb{E}_{x\sim q}dCor(G(F_t(x)), G'(F'_t(x)))$   

//disagreement restriction for shared information $G(F_t(x))$ and target-critical information $G'(F_t'(x))$.

            $E_{\text{en}} = E_{\text{en}_1} +E_{\text{en}_2}$
             //ensemble for agreement and disagreement information};
        \EndFor
    \end{algorithmic}
\end{algorithm}

In this work, we propose a method, called ADGKT, including agreement and disagreement mechanisms.
The agreement mechanism consists of GradVac, which alleviates gradient conflicts, and LogitNorm, which mitigates dominant gradients. Additionally, the disagreement mechanism includes a disagreement restriction and an ensemble approach. We utilize the disagreement restriction to obtain distinct and independent information, thereby promoting diversity. Furthermore, we introduce an ensemble of models to capture diverse aspects of the target features.




\subsection{Agreement}

During the training process, the update of shared parameter may encounter gradient conflict and dominant gradient. 
As shown in Fig. \ref{agreement}, to address gradient conflict, we implement the GradVac method, which adapts the gradients $g_s$ to $g_s'$ during training to reduce gradient conflicts. 
Additionally, to mitigate the dominance gradient, we apply LogitNorm to get updated logits $\hat{z_s}$ and $\hat{z_t}$, controlling their magnitude to prevent potential underfitting.

\subsubsection{Alleviating Gradient Conflict through GradVac}


To alleviate gradient conflict, we introduce GradVac, which adjusts the gradients between source and target during training to ensure they are more aligned and less likely to interfere with each other. First, the cosine similarity score $\text{cos}(\theta) = \phi$ for the gradients $g_s$ and $g_t$ can be defined:
\begin{equation}
   \text{cos}(\theta) = \frac{g_s \cdot g_t}{\|g_s\| \|g_t\|}
\end{equation}
where $\theta$ represents the angle between the two gradients. This similarity score $\phi$ helps to quantify the degree of conflict between the gradients. When $\phi < \alpha$, the gradient conflict becomes significant and cannot be ignored. In this case, the gradient $g_s$ is updated to a new gradient $g_s'$ to mitigate the conflict:
\begin{equation}
   g_s' = g_s + \eta g_t
\end{equation}
where $\eta$ is determined by the Law of Sines, calculated as:
\begin{equation}
   \eta = \frac{\|g_s\|(\phi^T \sqrt{1-\phi^2} - \phi \sqrt{1-(\phi^T)^2})}{\|g_t\| \sqrt{1-(\phi^T)^2}}
\end{equation}
The threshold $\alpha$ 
is dynamically updated at training step $t$ through an exponential moving average (EMA):
\begin{equation}
\label{ema}
\alpha^{(t)} =  (1-\beta)\alpha^{(t-1)} + \beta\phi^{(t-1)}
\end{equation}
where $\phi^{(t-1)}$ is the cosine similarity of the gradients $g_s$ and $g_t$ at training step $t-1$, $\beta$ is a hyperparameter, and $\alpha^{(0)}=0$.

\subsubsection{Mitigating the Dominating Gradients via LogitNorm}


To reduce the impact of dominating gradient, we employ the gradient magnitude similarity $\Phi(g_s, g_t)$ to represent the contribution between source gradient $g_s$ and target gradient $g_t$ for shared parameter. 
The gradient magnitude similarity \cite{yu2020gradient} can be represent as: 
\begin{equation}
    \Phi(g_s, g_t) = \frac{2 \|g_s\|_2 \|g_t\|_2}{\|g_s\|_2^2 + \|g_t\|_2^2}
\end{equation}
The value of $\Phi$ is between 0 and 1. When the two gradients $g_s$ and $g_t$ have the same magnitude ($\|g_s\|_2 = \|g_t\|_2 $), the similarity $\Phi$ equals 1, which means the source and target scenes have the positive impact on the shared parameter. When the magnitude of two gradients are significant different, $\Phi$ equals 0, leading to imbalanced optimizing for shared feature, thereby causing sub-optimizing the shared parameter.
In addition, the cross-entropy loss $L$ encourages the model to increase the magnitude of logits  $\|z\|$ during the training process.
The logit gradient of cross-entropy loss can be represented as:
\begin{equation}
    \frac{\partial L}{\partial z_k} = \hat{y}_k - y_k
\end{equation}
where $\hat{y}_k$ denotes predicted probability for class $k$. $y_k$ represents the true label for class $k$. 
When source scene contain more samples than target scene, source scene can have high frequency to update the shared parameter for source presentation. Over time, the magnitude of $\hat{y}_k$ for dominant source task will increase, thereby, the magnitude of source logits will grow larger than target.

To address the dominant gradient, we apply LogitNorm in Eq. \eqref{norm}. By controlling the gradient magnitudes, LogitNorm can better balance the importance of source and target scenes during the shared parameter optimization.
\begin{equation}
\label{norm}
\hat{{z}} = \frac{{z}}{\tau\|{z}\|}
\end{equation}
where $\hat{{z}}$ represents the normalized logits, $ \|{z}\| $ is the magnitude of the logits vector, and $\tau$ denotes the temperature, which regulates the magnitude of the logits.
The modified cross-entropy loss function with LogitNorm is:
\begin{equation}
\mathcal{L}(\hat{z}_i, y) = - \log \left( \frac{\exp(\hat{z}_i)}{\sum_{j=1}^{C} \exp(\hat{z}_j)} \right)
\end{equation}
\subsection{Disagreement}

\begin{figure}[t]
    \centering
    \includegraphics[width=\linewidth]{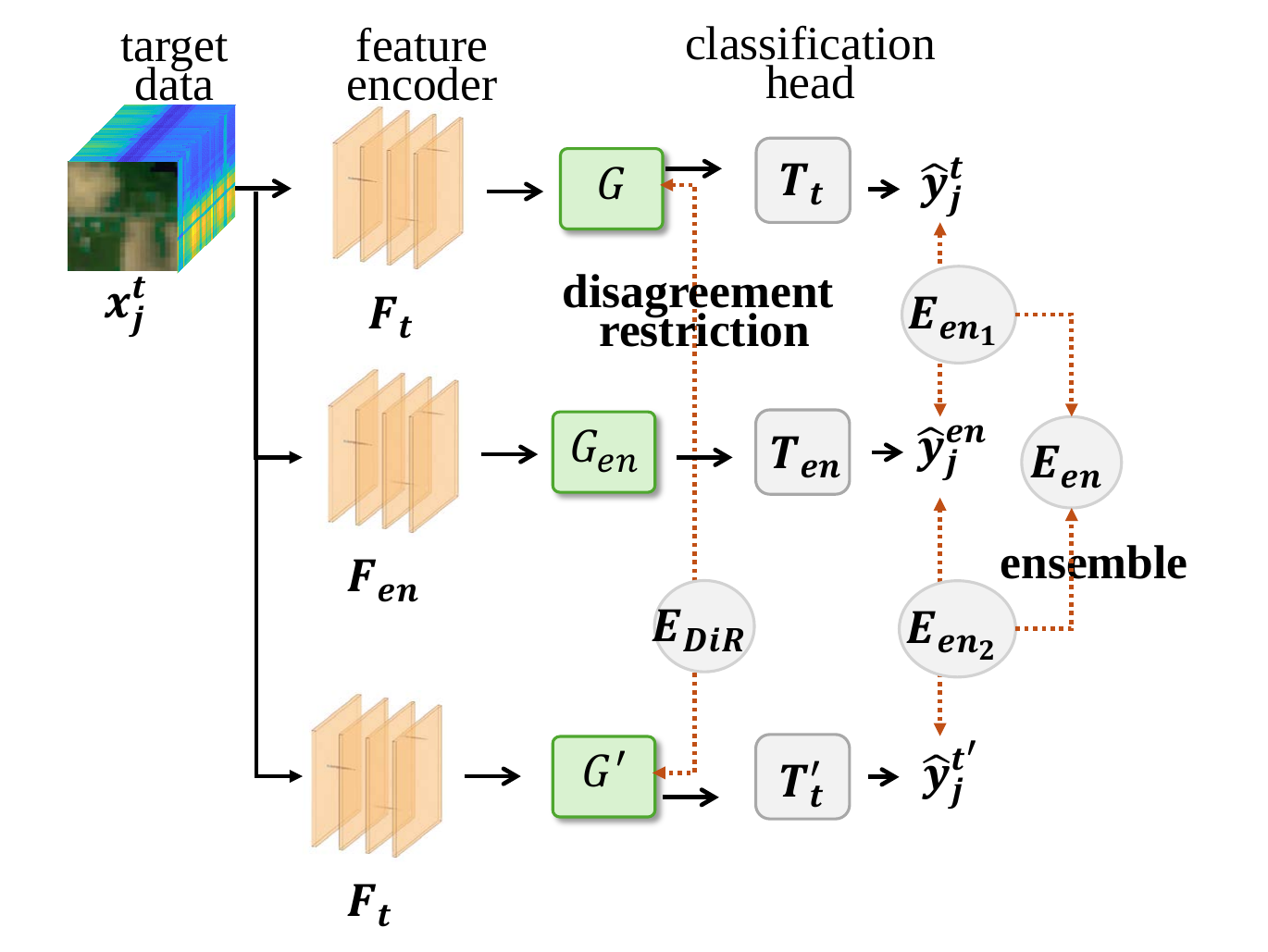}
    \caption{The disagreement mechanism is introduced in our model. To prevent the risk of losing critical target features, we employ disagreement restriction through $E_{\text{DiR}}$ to promote diversity. Then, we utilize the ensemble component through $E_{\text{en}}$ to capture a diverse set of target features.  }
    \label{disagreement}
\end{figure}
When focusing on achieving agreement between source and target scenes, there is a risk of missing the rich, diverse patterns present in the target scene. To prevent this, we introduce disagreement mechanism including a disagreement restriction using $E_{\text{DiR}}$ and an ensemble component using $E_{\text{en}}$ to promote target disagreement, as illustrated in Fig.\ref{disagreement}.

\subsubsection{Promoting diversity through disagreement restriction}
To enhance the predictive power of the target features during the training process, it is essential to promote diversity and independence among them. To achieve this, we employ a disagreement restriction (DiR) using partial distance correlation \cite{zhen2022versatile}, which enforces orthogonality among features:

\begin{equation}
\label{dc_target}
E_{\text{DiR}} = \mathbb{E}_{x\sim q}dCor(G(F_t(x)), G'(F'_t(x)))
\end{equation}
where $F_t', G'$ and $T_t'$ denote the separate components to capture target critical information $G'(F_t'(x))$, which is enforced to be orthogonal to the shared information between the source and target scenes, represented by $G(F_t(x))$.

By applying this disagreement restriction, each feature is encouraged to capture distinct and independent aspects of the target data. Therefore, the disagreement restriction promotes a diverse representation of the target scene, ensuring that important target critical information is effectively learned.

\subsubsection{Capturing a diverse set of target features through ensemble}

We propose an ensemble of models to capture a diverse set of features for the target scene. This ensemble approach enables the model to integrate multiple perspectives of the target features. Additionally, we employ reverse distillation \cite{li2022shadow} to iteratively reduce the discrepancy between the teacher models (which include both agreement and disagreement mechanisms) and the ensemble model, thereby updating and refining the integrated information through the ensemble.
\begin{equation}
\label{kt_1}
\begin{split}
E_{\text{en}_1} & = \mathbb{E}_{x\sim q} \big [D_{\text{KL}}(T_{en}(G_{en}(F_t(x))), T_t(G(F_t(x))))\\ 
&+ D_{\text{KL}}( T_t(G(F_t(x))), T_{en}(G_{en}(F_t(x)))) \big ]
\end{split}
\end{equation}

\begin{align}
E_{\text{en}_2} & = \mathbb{E}_{x\sim q} \big [ D_{\text{KL}}(T_{en}(G_{en}(F_t(x))), T'_t(G'(F_t'(x)))) \nonumber \\ 
&+ D_{\text{KL}}( T'_t(G'(F_t'(x))), T_{en}(G_{en}(F_t(x))))  \big ] \label{kt_2}
\end{align}
where $D_{\text{KL}}$ denotes KL divergence. 

The knowledge distillation (KD) loss for the ensemble model coupled with the agreement mechanisms is represented by $E_{\text{en}_1}$ in Eq.\eqref{kt_1}. Similarly, $E_{\text{en}_2}$ in Eq. \eqref{kt_2} represents the KD loss for the ensemble model with the disagreement mechanisms. By learning from the agreement and disagreement teacher, the final ensemble model can effectively integrate these two aspects of target information in Eq.\eqref{kl}, ensuring that model captures all essential features necessary for the target task.



\begin{equation}
\label{kl}
E_{\text{en}} = E_{\text{en}_1} +E_{\text{en}_2} 
\end{equation}




\begin{table}[tb]
\caption{The detailed categories information for Indian Pines, Pavia University and Houston 2013 datasets.}
\label{table:datasets_class}
\begin{tabular}{c|c|c}
\hline
 Indian Pines & Pavia University & Houston 2013 \\ \hline
 Alfalfa & Asphalt & Health grass  \\
  Corn-notill & Meadows & Stressed grass  \\
 Corn-mintill & Gravel & Synthetic grass\\
  Corn & Trees & Tress \\
 Grass-pasture & Painted metal sheets  & Soil \\
 Grass-trees & Bare Soil & Water \\
 Grass-pasture-mowed & Bitumen & Residential \\
 Hay-windrowed & Self-Blocking Bricks & Commercial \\
 Oats & Shadows & Road \\
 Soybean-notill &  & Highway \\
 Soybean-mintill &  & Railway \\
 Soybean-clean &  & Parking lot 1 \\
 Wheat &  & Parking lot 2 \\
 Woods &  & Tennis court\\
 Buildings-Grass-Trees-Drives &  & Running track  \\
 Stone-Steel-Towers &  &  \\ \hline
\end{tabular}
\end{table}

\begin{sidewaystable}[htbp]
\caption{The performance (\%) of knowledge transfer from IndianPine to Pavia (I$\rightarrow$ P) and from Houston to Pavia (H $\rightarrow$ P). $\uparrow$ indicates that higher values are better. \textbf{Bold} numbers represent the best results.}
\label{Pavia}
\setlength{\tabcolsep}{2pt}
\begin{tabular*}{\textheight}{@{\extracolsep{\fill}} cc|ccccccc|ccccccc}
\hline
\multirow{2}{*}{Class} & \multirow{2}{*}{Baseline}
  & \multicolumn{7}{c|}{I$\rightarrow$ P}
  & \multicolumn{7}{c}{H $\rightarrow$ P} \\ 
 & & MTL & UAN & ONE & FFL & Adaptor & Finetune & Ours
  & MTL & UAN & ONE & FFL & Adaptor & Finetune & Ours  \\ \hline
                 Asphalt &83.68& \multicolumn{1}{c|}{78.66} & \multicolumn{1}{c|}{78.11} & \multicolumn{1}{c|}{79.70} & \multicolumn{1}{c|}{73.65} & \multicolumn{1}{c|}{\textbf{85.42}} & \multicolumn{1}{c|}{79.74} & {86.45}  & \multicolumn{1}{c|}{77.62} & \multicolumn{1}{c|}{78.78} & \multicolumn{1}{c|}{87.06} & \multicolumn{1}{c|}{85.03} & \multicolumn{1}{c|}{79.17} & \multicolumn{1}{c|}{{83.82}} & \textbf{88.59}  \\ \hline
                 Meadows& 69.02 & \multicolumn{1}{c|}{69.50} & \multicolumn{1}{c|}{82.34} & \multicolumn{1}{c|}{63.72} & \multicolumn{1}{c|}{73.39} & \multicolumn{1}{c|}{81.57} & \multicolumn{1}{c|}{{86.10}} & \textbf{94.44} & \multicolumn{1}{c|}{59.80} & \multicolumn{1}{c|}{74.50} & \multicolumn{1}{c|}{71.46} & \multicolumn{1}{c|}{68.47} & \multicolumn{1}{c|}{76.64} & \multicolumn{1}{c|}{{80.31}} & \textbf{92.29}  \\ \hline
                 Gravel & 57.25 & \multicolumn{1}{c|}{73.72} & \multicolumn{1}{c|}{66.89} & \multicolumn{1}{c|}{73.28} & \multicolumn{1}{c|}{71.46} & \multicolumn{1}{c|}{\textbf{87.99}} & \multicolumn{1}{c|}{68.32} & {76.64} & \multicolumn{1}{c|}{74.93} & \multicolumn{1}{c|}{73.22} & \multicolumn{1}{c|}{71.35} & \multicolumn{1}{c|}{67.27} & \multicolumn{1}{c|}{73.39} & \multicolumn{1}{c|}{63.47} & {\textbf{75.87}} \\ \hline
                Trees& 90.01  & \multicolumn{1}{c|}{81.73} & \multicolumn{1}{c|}{82.49} & \multicolumn{1}{c|}{84.51} & \multicolumn{1}{c|}{\textbf{91.14}} & \multicolumn{1}{c|}{86.44} & \multicolumn{1}{c|}{{87.81}} & 84.68 & \multicolumn{1}{c|}{87.23} & \multicolumn{1}{c|}{86.47} & \multicolumn{1}{c|}{{90.18}} & \multicolumn{1}{c|}{88.32} & \multicolumn{1}{c|}{\textbf{94.71}} & \multicolumn{1}{c|}{82.66} & 84.10  \\ \hline
                Painted metal sheets& \textbf{99.01}  & \multicolumn{1}{c|}{94.34} & \multicolumn{1}{c|}{96.32} & \multicolumn{1}{c|}{87.78} & \multicolumn{1}{c|}{97.21} & \multicolumn{1}{c|}{96.77} & \multicolumn{1}{c|}{\textbf{99.10}} & {9605} & \multicolumn{1}{c|}{96.86} & \multicolumn{1}{c|}{95.87} & \multicolumn{1}{c|}{{98.38}} & \multicolumn{1}{c|}{97.12} & \multicolumn{1}{c|}{{98.65}} & \multicolumn{1}{c|}{90.57} &  93.62  \\ \hline
                 Bare Soil & 55.51 & \multicolumn{1}{c|}{\textbf{79.31}} & \multicolumn{1}{c|}{69.79} & \multicolumn{1}{c|}{{85.74}} & \multicolumn{1}{c|}{50.79} & \multicolumn{1}{c|}{79.18} & \multicolumn{1}{c|}{65.46} & {64.52}& \multicolumn{1}{c|}{\textbf{91.84}} & \multicolumn{1}{c|}{55.03} & \multicolumn{1}{c|}{54.64} & \multicolumn{1}{c|}{67.98} & \multicolumn{1}{c|}{41.23} & \multicolumn{1}{c|}{61.70} & {61.61} \\ \hline
                 Bitumen& 68.50 & \multicolumn{1}{c|}{70.03} & \multicolumn{1}{c|}{69.42} & \multicolumn{1}{c|}{58.00} & \multicolumn{1}{c|}{{72.17}} & \multicolumn{1}{c|}{56.57} & \multicolumn{1}{c|}{68.60} & \textbf{64.63} & \multicolumn{1}{c|}{\textbf{75.74}} & \multicolumn{1}{c|}{74.21} & \multicolumn{1}{c|}{67.99} & \multicolumn{1}{c|}{60.35} & \multicolumn{1}{c|}{69.22} & \multicolumn{1}{c|}{64.93} & { 69.42}  \\ \hline
                Self-Blocking Bricks& \textbf{95.04}  & \multicolumn{1}{c|}{93.61} & \multicolumn{1}{c|}{\textbf{95.30}} & \multicolumn{1}{c|}{89.80} & \multicolumn{1}{c|}{\textbf{95.30}} & \multicolumn{1}{c|}{86.95} & \multicolumn{1}{c|}{93.01} & {94.20} & \multicolumn{1}{c|}{91.08} & \multicolumn{1}{c|}{92.24} & \multicolumn{1}{c|}{{93.64}} & \multicolumn{1}{c|}{{94.44}} & \multicolumn{1}{c|}{83.44} & \multicolumn{1}{c|}{93.82} &  94.83  \\ \hline
                 Shadows & 93.08 & \multicolumn{1}{c|}{88.55} & \multicolumn{1}{c|}{85.41} & \multicolumn{1}{c|}{87.42} & \multicolumn{1}{c|}{{91.57}} & \multicolumn{1}{c|}{89.43} & \multicolumn{1}{c|}{91.19} & {\textbf{93.71}} & \multicolumn{1}{c|}{89.94} & \multicolumn{1}{c|}{89.43} & \multicolumn{1}{c|}{92.33} & \multicolumn{1}{c|}{{93.71}} & \multicolumn{1}{c|}{91.45} & \multicolumn{1}{c|}{\textbf{94.34}} & 90.57 \\ \hline
                OA $\uparrow$ & 74.27 & \multicolumn{1}{c|}{76.26} & \multicolumn{1}{c|}{80.77} & \multicolumn{1}{c|}{73.90} & \multicolumn{1}{c|}{74.89} & \multicolumn{1}{c|}{{82.97}} & \multicolumn{1}{c|}{82.67} & \textbf{87.52} & \multicolumn{1}{c|}{73.61} & \multicolumn{1}{c|}{76.14} & \multicolumn{1}{c|}{76.30} & \multicolumn{1}{c|}{75.70} & \multicolumn{1}{c|}{75.46} & \multicolumn{1}{c|}{{79.69}} & \textbf{82.88}  \\ \hline
                AA $\uparrow$  & 79.01 & \multicolumn{1}{c|}{81.05} & \multicolumn{1}{c|}{80.67} & \multicolumn{1}{c|}{78.88} & \multicolumn{1}{c|}{79.63} & \multicolumn{1}{c|}{83.37} & \multicolumn{1}{c|}{82.15} & {\textbf{83.93}}  & \multicolumn{1}{c|}{\textbf{82.78}} & \multicolumn{1}{c|}{79.97} & \multicolumn{1}{c|}{80.78} & \multicolumn{1}{c|}{80.30} & \multicolumn{1}{c|}{78.66} & \multicolumn{1}{c|}{81.09} & {80.29} \\ \hline
               $\kappa$ $\uparrow$  & 66.93  & \multicolumn{1}{c|}{69.62} & \multicolumn{1}{c|}{74.61} & \multicolumn{1}{c|}{66.91} & \multicolumn{1}{c|}{67.52} & \multicolumn{1}{c|}{{77.61}} & \multicolumn{1}{c|}{76.91} & \textbf{83.01} & \multicolumn{1}{c|}{67.11} & \multicolumn{1}{c|}{68.93} & \multicolumn{1}{c|}{69.40} & \multicolumn{1}{c|}{68.89} & \multicolumn{1}{c|}{67.85} & \multicolumn{1}{c|}{{73.17}} & \textbf{77.03}\\ \hline
\end{tabular*}
\end{sidewaystable}


\begin{sidewaystable}[htbp]
\centering
\caption{The performance (\%) of knowledge transfer from from Pavia to Houston (P $\rightarrow$ H) and IndianPine to Houston (I$\rightarrow$ H). $\uparrow$ indicates that higher values are better. \textbf{Bold} numbers represent the best results. }
\label{houston}
\setlength{\tabcolsep}{2pt}
\begin{tabular*}{\textheight}{@{\extracolsep{\fill}} cc|ccccccc|ccccccc}
\hline
\multirow{2}{*}{Class} & \multirow{2}{*}{Baseline}
  & \multicolumn{7}{c|}{P$\rightarrow$ H}
  & \multicolumn{7}{c}{I$\rightarrow$ H} \\ 
 & & MTL & UAN & ONE & FFL & Adaptor & Finetune & Ours
  & MTL & UAN & ONE & FFL & Adaptor & Finetune & Ours  \\ \hline
   Healthy grass& 74.55 & \multicolumn{1}{c|}{70.28} & \multicolumn{1}{c|}{68.85} & \multicolumn{1}{c|}{63.53} & \multicolumn{1}{c|}{70.66} & \multicolumn{1}{c|}{74.55} & \multicolumn{1}{c|}{76.26} & \textbf{78.44} & \multicolumn{1}{c|}{75.59} & \multicolumn{1}{c|}{71.89} & \multicolumn{1}{c|}{76.92} & \multicolumn{1}{c|}{76.35} & \multicolumn{1}{c|}{69.99} & \multicolumn{1}{c|}{73.03} & \textbf{80.25}  \\ \hline
                 Stressed grass& 89.85 & \multicolumn{1}{c|}{92.67} & \multicolumn{1}{c|}{95.68} & \multicolumn{1}{c|}{97.37} & \multicolumn{1}{c|}{97.27} & \multicolumn{1}{c|}{94.83} & \multicolumn{1}{c|}{92.95} & \textbf{97.84} & \multicolumn{1}{c|}{\textbf{95.86}} & \multicolumn{1}{c|}{95.21} & \multicolumn{1}{c|}{96.90} & \multicolumn{1}{c|}{97.65} & \multicolumn{1}{c|}{95.39} & \multicolumn{1}{c|}{92.01} & 95.58  \\ \hline
                 Synthetic grass& 54.65 & \multicolumn{1}{c|}{70.89} & \multicolumn{1}{c|}{\textbf{75.25}} & \multicolumn{1}{c|}{74.65} & \multicolumn{1}{c|}{68.32} & \multicolumn{1}{c|}{73.27} & \multicolumn{1}{c|}{48.71} &  69.50 & \multicolumn{1}{c|}{60.20} & \multicolumn{1}{c|}{64.75} & \multicolumn{1}{c|}{74.06} & \multicolumn{1}{c|}{79.80} & \multicolumn{1}{c|}{75.64} & \multicolumn{1}{c|}{\textbf{80.79}} & 72.48 \\ \hline
                 Trees&91.66  & \multicolumn{1}{c|}{\textbf{97.06}} & \multicolumn{1}{c|}{91.57} & \multicolumn{1}{c|}{96.31} & \multicolumn{1}{c|}{96.88} & \multicolumn{1}{c|}{93.66} & \multicolumn{1}{c|}{96.02} &  96.78 & \multicolumn{1}{c|}{89.20} & \multicolumn{1}{c|}{96.12} & \multicolumn{1}{c|}{97.25} & \multicolumn{1}{c|}{96.59} & \multicolumn{1}{c|}{95.74} & \multicolumn{1}{c|}{88.73} & \textbf{98.39}  \\ \hline
                Soil& 99.53  & \multicolumn{1}{c|}{99.34} & \multicolumn{1}{c|}{97.82} & \multicolumn{1}{c|}{99.43} & \multicolumn{1}{c|}{96.31} & \multicolumn{1}{c|}{98.20} & \multicolumn{1}{c|}{\textbf{99.62}} & 99.24 & \multicolumn{1}{c|}{99.43} & \multicolumn{1}{c|}{\textbf{100.00}} & \multicolumn{1}{c|}{98.86} & \multicolumn{1}{c|}{99.15} & \multicolumn{1}{c|}{98.11} & \multicolumn{1}{c|}{\textbf{99.81}} & 99.72  \\ \hline
                 Water & 77.62 & \multicolumn{1}{c|}{74.13} & \multicolumn{1}{c|}{81.82} & \multicolumn{1}{c|}{81.82} & \multicolumn{1}{c|}{\textbf{82.52}} & \multicolumn{1}{c|}{69.23} & \multicolumn{1}{c|}{70.63} & 81.82 & \multicolumn{1}{c|}{\textbf{83.22}} & \multicolumn{1}{c|}{77.62} & \multicolumn{1}{c|}{81.12} & \multicolumn{1}{c|}{80.42} & \multicolumn{1}{c|}{69.23} & \multicolumn{1}{c|}{65.03} &  80.42 \\ \hline
                Residential& 78.82  & \multicolumn{1}{c|}{88.25} & \multicolumn{1}{c|}{86.75} & \multicolumn{1}{c|}{79.48} & \multicolumn{1}{c|}{86.01} & \multicolumn{1}{c|}{84.24} & \multicolumn{1}{c|}{86.38} &  \textbf{88.62}  & \multicolumn{1}{c|}{84.24} & \multicolumn{1}{c|}{83.77} & \multicolumn{1}{c|}{83.02} & \multicolumn{1}{c|}{83.86} & \multicolumn{1}{c|}{77.71} & \multicolumn{1}{c|}{72.20} & \textbf{89.09} \\ \hline
                Commercial & \textbf{83.38}   & \multicolumn{1}{c|}{68.76} & \multicolumn{1}{c|}{68.28} & \multicolumn{1}{c|}{72.27} & \multicolumn{1}{c|}{73.12} & \multicolumn{1}{c|}{68.38} & \multicolumn{1}{c|}{76.07} & 81.39 & \multicolumn{1}{c|}{77.59} & \multicolumn{1}{c|}{78.44} & \multicolumn{1}{c|}{76.07} & \multicolumn{1}{c|}{68.57} & \multicolumn{1}{c|}{\textbf{85.28}} & \multicolumn{1}{c|}{87.27} & 73.41\\ \hline   
                Road & 82.34 & \multicolumn{1}{c|}{71.01} & \multicolumn{1}{c|}{69.78} & \multicolumn{1}{c|}{\textbf{84.23}} & \multicolumn{1}{c|}{62.04} & \multicolumn{1}{c|}{82.91} & \multicolumn{1}{c|}{66.57} & 77.24 & \multicolumn{1}{c|}{63.36} & \multicolumn{1}{c|}{64.78} & \multicolumn{1}{c|}{79.23} & \multicolumn{1}{c|}{67.61} & \multicolumn{1}{c|}{\textbf{86.87}} & \multicolumn{1}{c|}{78.94} & 77.05 \\ \hline       
                 Highway& 62.36 & \multicolumn{1}{c|}{72.49} & \multicolumn{1}{c|}{55.21} & \multicolumn{1}{c|}{56.56} & \multicolumn{1}{c|}{68.82} & \multicolumn{1}{c|}{48.84} & \multicolumn{1}{c|}{53.47} &  \textbf{77.12} & \multicolumn{1}{c|}{51.74} & \multicolumn{1}{c|}{66.51} & \multicolumn{1}{c|}{51.74} & \multicolumn{1}{c|}{66.12} & \multicolumn{1}{c|}{62.93} & \multicolumn{1}{c|}{55.50} &  \textbf{83.98}\\ \hline
                Railway& 70.97  & \multicolumn{1}{c|}{77.80} & \multicolumn{1}{c|}{80.74} & \multicolumn{1}{c|}{82.07} & \multicolumn{1}{c|}{82.45} & \multicolumn{1}{c|}{80.36} & \multicolumn{1}{c|}{\textbf{86.81}} & 78.93 & \multicolumn{1}{c|}{73.72} & \multicolumn{1}{c|}{80.83} & \multicolumn{1}{c|}{\textbf{86.15}} & \multicolumn{1}{c|}{78.08} & \multicolumn{1}{c|}{79.32} & \multicolumn{1}{c|}{78.56} & 79.89  \\ \hline
                 Parking Lot 1& 71.47 & \multicolumn{1}{c|}{71.18} & \multicolumn{1}{c|}{72.43} & \multicolumn{1}{c|}{64.07} & \multicolumn{1}{c|}{62.34} & \multicolumn{1}{c|}{70.32} & \multicolumn{1}{c|}{57.44} & \textbf{72.62} & \multicolumn{1}{c|}{\textbf{73.20}} & \multicolumn{1}{c|}{68.20} & \multicolumn{1}{c|}{68.78} & \multicolumn{1}{c|}{66.57} & \multicolumn{1}{c|}{67.72} & \multicolumn{1}{c|}{58.50} & 69.84  \\ \hline
                 Parking Lot 2 & 90.18 & \multicolumn{1}{c|}{87.37} & \multicolumn{1}{c|}{88.77} & \multicolumn{1}{c|}{86.67} & \multicolumn{1}{c|}{82.81} & \multicolumn{1}{c|}{78.95} & \multicolumn{1}{c|}{81.75} & \textbf{90.18} & \multicolumn{1}{c|}{84.21} & \multicolumn{1}{c|}{84.21} & \multicolumn{1}{c|}{85.96} & \multicolumn{1}{c|}{89.47} & \multicolumn{1}{c|}{72.63} & \multicolumn{1}{c|}{84.56} & \textbf{92.28}\\ \hline
               Tennis Court  & 81.38   & \multicolumn{1}{c|}{81.78} & \multicolumn{1}{c|}{90.28} & \multicolumn{1}{c|}{78.54} & \multicolumn{1}{c|}{\textbf{92.71}} & \multicolumn{1}{c|}{83.40} & \multicolumn{1}{c|}{76.11} &  91.90 & \multicolumn{1}{c|}{\textbf{90.28}} & \multicolumn{1}{c|}{88.07} & \multicolumn{1}{c|}{91.09} & \multicolumn{1}{c|}{87.04} & \multicolumn{1}{c|}{73.68} & \multicolumn{1}{c|}{88.26} & 86.64\\ \hline
               Running Track  & 69.13  & \multicolumn{1}{c|}{64.27} & \multicolumn{1}{c|}{67.02} & \multicolumn{1}{c|}{75.48} & \multicolumn{1}{c|}{78.01} & \multicolumn{1}{c|}{57.29} & \multicolumn{1}{c|}{\textbf{84.14}} & 79.70 & \multicolumn{1}{c|}{65.75} & \multicolumn{1}{c|}{71.88} & \multicolumn{1}{c|}{\textbf{86.05}} & \multicolumn{1}{c|}{79.70} & \multicolumn{1}{c|}{65.33} & \multicolumn{1}{c|}{54.55} & 73.36 \\ \hline
               OA ($\uparrow$)  & 79.24  & \multicolumn{1}{c|}{79.96} & \multicolumn{1}{c|}{78.69} & \multicolumn{1}{c|}{79.42} & \multicolumn{1}{c|}{79.49} & \multicolumn{1}{c|}{78.52} & \multicolumn{1}{c|}{78.08} & \textbf{84.27}& \multicolumn{1}{c|}{77.65} & \multicolumn{1}{c|}{79.85} & \multicolumn{1}{c|}{81.73} & \multicolumn{1}{c|}{80.45} & \multicolumn{1}{c|}{80.53} & \multicolumn{1}{c|}{77.87} & \textbf{83.97} \\ \hline
                AA ($\uparrow$) & 78.53  & \multicolumn{1}{c|}{79.15} & \multicolumn{1}{c|}{79.35} & \multicolumn{1}{c|}{79.50} & \multicolumn{1}{c|}{80.02} & \multicolumn{1}{c|}{77.23} & \multicolumn{1}{c|}{76.86} & \textbf{84.09} & \multicolumn{1}{c|}{77.84} & \multicolumn{1}{c|}{79.49} & \multicolumn{1}{c|}{82.21} & \multicolumn{1}{c|}{81.13} & \multicolumn{1}{c|}{78.37} & \multicolumn{1}{c|}{77.18} & \textbf{83.49} \\ \hline
               $\kappa$ ($\uparrow$) & 77.52  & \multicolumn{1}{c|}{78.26} & \multicolumn{1}{c|}{76.88} & \multicolumn{1}{c|}{77.67} & \multicolumn{1}{c|}{77.76} & \multicolumn{1}{c|}{76.69} & \multicolumn{1}{c|}{76.21} & \textbf{82.92} & \multicolumn{1}{c|}{75.77} & \multicolumn{1}{c|}{78.15} & \multicolumn{1}{c|}{80.18} & \multicolumn{1}{c|}{78.80} & \multicolumn{1}{c|}{78.84} & \multicolumn{1}{c|}{75.98} & \textbf{82.61}  \\ 
\hline
\end{tabular*}
\end{sidewaystable}

\begin{sidewaystable}[htbp]
\centering
\caption{The performance (\%) of knowledge transfer from Pavia to IndianPine  (P$\rightarrow$ I) and from Houston to IndianPian (H $\rightarrow$ I). $\uparrow$ indicates that higher values are better. \textbf{Bold} numbers represent the best results.}
\label{IndianPine}
\setlength{\tabcolsep}{0.6pt}
\begin{tabular*}{\textheight}{@{\extracolsep{\fill}} cc|ccccccc|ccccccc}
\hline
\multirow{2}{*}{Class} & \multirow{2}{*}{Baseline}
  & \multicolumn{7}{c|}{P$\rightarrow$ I}
  & \multicolumn{7}{c}{H$\rightarrow$ I} \\ 
 & & MTL & UAN & ONE & FFL & Adaptor & Finetune & Ours
  & MTL & UAN & ONE & FFL & Adaptor & Finetune & Ours  \\ \hline 

                 Alfalfa& 66.04 & \multicolumn{1}{c|}{59.10} & \multicolumn{1}{c|}{56.29} & \multicolumn{1}{c|}{61.42} & \multicolumn{1}{c|}{66.84} & \multicolumn{1}{c|}{\textbf{68.21}} & \multicolumn{1}{c|}{58.09} & 62.28 & \multicolumn{1}{c|}{55.63} & \multicolumn{1}{c|}{48.77} & \multicolumn{1}{c|}{62.21} & \multicolumn{1}{c|}{58.53} & \multicolumn{1}{c|}{61.27} & \multicolumn{1}{c|}{64.09} & \textbf{77.46} \\ \hline
                Corn-notill& 89.29  & \multicolumn{1}{c|}{69.26} & \multicolumn{1}{c|}{73.09} & \multicolumn{1}{c|}{83.93} & \multicolumn{1}{c|}{80.10} & \multicolumn{1}{c|}{61.73} & \multicolumn{1}{c|}{76.53} & \textbf{90.82}& \multicolumn{1}{c|}{68.11} & \multicolumn{1}{c|}{65.69} & \multicolumn{1}{c|}{83.93} & \multicolumn{1}{c|}{83.29} & \multicolumn{1}{c|}{83.16} & \multicolumn{1}{c|}{72.70} & \textbf{88.65} \\ \hline
                 Corn-mintill& \textbf{99.46} & \multicolumn{1}{c|}{86.96} & \multicolumn{1}{c|}{89.67} & \multicolumn{1}{c|}{93.48} & \multicolumn{1}{c|}{94.02} & \multicolumn{1}{c|}{87.50} & \multicolumn{1}{c|}{92.39} & 92.93 & \multicolumn{1}{c|}{95.65} & \multicolumn{1}{c|}{89.13} & \multicolumn{1}{c|}{93.48} & \multicolumn{1}{c|}{95.11} & \multicolumn{1}{c|}{91.30} & \multicolumn{1}{c|}{90.22} & 93.48 \\ \hline
                 
                Corn& \textbf{85.01} & \multicolumn{1}{c|}{83.67} & \multicolumn{1}{c|}{73.15} & \multicolumn{1}{c|}{84.34} & \multicolumn{1}{c|}{79.87} & \multicolumn{1}{c|}{78.75} & \multicolumn{1}{c|}{79.19} & \textbf{87.47} & \multicolumn{1}{c|}{83.45} & \multicolumn{1}{c|}{78.97} & \multicolumn{1}{c|}{78.30} & \multicolumn{1}{c|}{82.77} & \multicolumn{1}{c|}{81.43} & \multicolumn{1}{c|}{82.33} & 84.56  \\ \hline
                Grass-pasture& \textbf{91.39}& \multicolumn{1}{c|}{86.94} & \multicolumn{1}{c|}{78.05} & \multicolumn{1}{c|}{83.36} & \multicolumn{1}{c|}{88.67} & \multicolumn{1}{c|}{78.62} & \multicolumn{1}{c|}{78.77} &  86.80 & \multicolumn{1}{c|}{79.20} & \multicolumn{1}{c|}{66.86} & \multicolumn{1}{c|}{85.08} & \multicolumn{1}{c|}{79.34} & \multicolumn{1}{c|}{80.77} & \multicolumn{1}{c|}{75.61} & \textbf{91.68}  \\ \hline
                 Grass-trees& 93.62 & \multicolumn{1}{c|}{94.76} & \multicolumn{1}{c|}{94.76} & \multicolumn{1}{c|}{92.48} & \multicolumn{1}{c|}{93.62} & \multicolumn{1}{c|}{91.11} & \multicolumn{1}{c|}{92.26} & \textbf{94.99} & \multicolumn{1}{c|}{94.31} & \multicolumn{1}{c|}{94.08} & \multicolumn{1}{c|}{92.71} & \multicolumn{1}{c|}{94.08} & \multicolumn{1}{c|}{93.17} & \multicolumn{1}{c|}{90.66} & \textbf{94.99}  \\ \hline 
                 Grass-pasture-mowed& 70.26 & \multicolumn{1}{c|}{60.46} & \multicolumn{1}{c|}{54.58} & \multicolumn{1}{c|}{49.89} & \multicolumn{1}{c|}{68.52} & \multicolumn{1}{c|}{62.96} & \multicolumn{1}{c|}{62.31} & \textbf{75.16} & \multicolumn{1}{c|}{63.40} & \multicolumn{1}{c|}{66.56} & \multicolumn{1}{c|}{66.56} & \multicolumn{1}{c|}{61.33} & \multicolumn{1}{c|}{69.93} & \multicolumn{1}{c|}{55.12} &  \textbf{70.48}  \\ \hline
                 Hay-windrowed& 64.19  & \multicolumn{1}{c|}{58.60} & \multicolumn{1}{c|}{33.46} & \multicolumn{1}{c|}{57.73} & \multicolumn{1}{c|}{64.31} & \multicolumn{1}{c|}{68.78} & \multicolumn{1}{c|}{65.34} &  \textbf{72.66}  & \multicolumn{1}{c|}{50.95} & \multicolumn{1}{c|}{41.52} & \multicolumn{1}{c|}{56.99} & \multicolumn{1}{c|}{60.26} & \multicolumn{1}{c|}{66.67} & \multicolumn{1}{c|}{65.34} & \textbf{72.37} \\ \hline
                 Oats& \textbf{76.06} & \multicolumn{1}{c|}{70.39} & \multicolumn{1}{c|}{55.67} & \multicolumn{1}{c|}{68.44} & \multicolumn{1}{c|}{64.89} & \multicolumn{1}{c|}{67.02} & \multicolumn{1}{c|}{66.31} &  75.70 & \multicolumn{1}{c|}{66.31} & \multicolumn{1}{c|}{71.45} & \multicolumn{1}{c|}{68.44} & \multicolumn{1}{c|}{74.65} & \multicolumn{1}{c|}{47.70} & \multicolumn{1}{c|}{59.22} & 75.71 \\ \hline  
                  Soybean-notill& 95.68& \multicolumn{1}{c|}{\textbf{98.15}} & \multicolumn{1}{c|}{96.91} & \multicolumn{1}{c|}{97.53} & \multicolumn{1}{c|}{95.06} & \multicolumn{1}{c|}{96.91} & \multicolumn{1}{c|}{95.06} &  93.72 & \multicolumn{1}{c|}{94.44} & \multicolumn{1}{c|}{91.98} & \multicolumn{1}{c|}{95.68} & \multicolumn{1}{c|}{\textbf{98.77}} & \multicolumn{1}{c|}{88.27} & \multicolumn{1}{c|}{97.53} & 96.91 \\ \hline
                  Soybean-mintill& \textbf{93.89}& \multicolumn{1}{c|}{94.21} & \multicolumn{1}{c|}{92.12} & \multicolumn{1}{c|}{81.91} & \multicolumn{1}{c|}{90.19} & \multicolumn{1}{c|}{84.81} & \multicolumn{1}{c|}{87.86} & \textbf{97.78} & \multicolumn{1}{c|}{81.83} & \multicolumn{1}{c|}{83.04} & \multicolumn{1}{c|}{81.03} & \multicolumn{1}{c|}{93.17} & \multicolumn{1}{c|}{83.52} & \multicolumn{1}{c|}{91.80} & 93.49  \\ \hline
                 Soybean-clean& \textbf{87.88} & \multicolumn{1}{c|}{64.85} & \multicolumn{1}{c|}{72.12} & \multicolumn{1}{c|}{80.91} & \multicolumn{1}{c|}{74.24} & \multicolumn{1}{c|}{87.27} & \multicolumn{1}{c|}{78.79} &  \textbf{97.78}& \multicolumn{1}{c|}{77.58} & \multicolumn{1}{c|}{77.58} & \multicolumn{1}{c|}{78.48} & \multicolumn{1}{c|}{81.52} & \multicolumn{1}{c|}{83.94} & \multicolumn{1}{c|}{72.42} & 84.24 \\ \hline
                 Wheat & 95.56 & \multicolumn{1}{c|}{\textbf{100.00}} & \multicolumn{1}{c|}{\textbf{100.00}} & \multicolumn{1}{c|}{\textbf{100.00}} & \multicolumn{1}{c|}{\textbf{100.00}} & \multicolumn{1}{c|}{\textbf{100.00}} & \multicolumn{1}{c|}{\textbf{100.00}} & 97.78 & \multicolumn{1}{c|}{\textbf{100.00}} & \multicolumn{1}{c|}{97.78} & \multicolumn{1}{c|}{\textbf{100.00}} & \multicolumn{1}{c|}{\textbf{100.00}} & \multicolumn{1}{c|}{93.33} & \multicolumn{1}{c|}{\textbf{100.00}} & \textbf{100.00} \\ \hline
                 Woods & 89.74& \multicolumn{1}{c|}{97.44} & \multicolumn{1}{c|}{97.44} & \multicolumn{1}{c|}{94.87} & \multicolumn{1}{c|}{\textbf{100.00}} & \multicolumn{1}{c|}{\textbf{100.00}} & \multicolumn{1}{c|}{97.44} & \textbf{100.00} & \multicolumn{1}{c|}{\textbf{100.00}} & \multicolumn{1}{c|}{97.44} & \multicolumn{1}{c|}{\textbf{100.00}} & \multicolumn{1}{c|}{\textbf{100.00}} & \multicolumn{1}{c|}{\textbf{100.00}} & \multicolumn{1}{c|}{89.74} & \textbf{100.00} \\ \hline
                Buildings-Grass-Trees-Drives&\textbf{100.00}   & \multicolumn{1}{c|}{\textbf{100.00}} & \multicolumn{1}{c|}{\textbf{100.00}} & \multicolumn{1}{c|}{\textbf{100.00}} & \multicolumn{1}{c|}{\textbf{100.00}} & \multicolumn{1}{c|}{\textbf{100.00}} & \multicolumn{1}{c|}{\textbf{100.00}} & \textbf{100.00} & \multicolumn{1}{c|}{\textbf{100.00}} & \multicolumn{1}{c|}{\textbf{100.00}} & \multicolumn{1}{c|}{\textbf{100.00}} & \multicolumn{1}{c|}{\textbf{100.00}} & \multicolumn{1}{c|}{\textbf{100.00}} & \multicolumn{1}{c|}{\textbf{100.00}} &\textbf{100.00}  \\ \hline
                 Stone-Steel-Towers&\textbf{100.00}  & \multicolumn{1}{c|}{\textbf{100.00}} & \multicolumn{1}{c|}{\textbf{100.00}} & \multicolumn{1}{c|}{\textbf{100.00}} & \multicolumn{1}{c|}{\textbf{100.00}} & \multicolumn{1}{c|}{\textbf{100.00}} & \multicolumn{1}{c|}{\textbf{100.00}} & \textbf{100.00} & \multicolumn{1}{c|}{\textbf{100.00}} & \multicolumn{1}{c|}{\textbf{100.00}} & \multicolumn{1}{c|}{\textbf{100.00}} & \multicolumn{1}{c|}{\textbf{100.00}} & \multicolumn{1}{c|}{\textbf{100.00}} & \multicolumn{1}{c|}{\textbf{100.00}} &\textbf{100.00} \\ \hline
                OA ($\uparrow$) & 78.15 & \multicolumn{1}{c|}{71.66} & \multicolumn{1}{c|}{62.74} & \multicolumn{1}{c|}{70.58} & \multicolumn{1}{c|}{75.31} & \multicolumn{1}{c|}{73.50} & \multicolumn{1}{c|}{72.53} & \textbf{80.11} & \multicolumn{1}{c|}{67.56} & \multicolumn{1}{c|}{63.49} & \multicolumn{1}{c|}{71.78} & \multicolumn{1}{c|}{73.45} & \multicolumn{1}{c|}{73.24} & \multicolumn{1}{c|}{72.08} & \textbf{81.64}   \\ \hline
                AA ($\uparrow$)& 87.38  & \multicolumn{1}{c|}{ 82.80} & \multicolumn{1}{c|}{79.21} & \multicolumn{1}{c|}{83.14} & \multicolumn{1}{c|}{85.02} & \multicolumn{1}{c|}{83.36} & \multicolumn{1}{c|}{83.15} & \textbf{88.46} & \multicolumn{1}{c|}{81.93} & \multicolumn{1}{c|}{79.43} & \multicolumn{1}{c|}{84.08} & \multicolumn{1}{c|}{85.17} & \multicolumn{1}{c|}{82.78} & \multicolumn{1}{c|}{81.67} & \textbf{89.00} \\ \hline
                $\kappa$ ($\uparrow$) & 75.30 & \multicolumn{1}{c|}{68.12} & \multicolumn{1}{c|}{58.62} & \multicolumn{1}{c|}{66.90} & \multicolumn{1}{c|}{72.06} & \multicolumn{1}{c|}{69.84} & \multicolumn{1}{c|}{68.92} & \textbf{77.39} & \multicolumn{1}{c|}{63.79} & \multicolumn{1}{c|}{59.41} & \multicolumn{1}{c|}{68.28} & \multicolumn{1}{c|}{70.17} & \multicolumn{1}{c|}{69.65} & \multicolumn{1}{c|}{68.23} &  \textbf{79.10}  \\ \hline
\end{tabular*}
\end{sidewaystable}

\section{Experiments}

\subsection{Experimental Setup}
We conduct experiments on three popular datasets: Indian Pines (I), Pavia (P), and Houston2013 (H). 
There is no straight forward categories correspondence for these three datasets, as shown in Table. \ref{table:datasets_class}. In addition, these three datasets captured by distinct HSI sensors.
During the knowledge transfer process, all training samples from the source scenes are utilized, while only 10 randomly selected samples for each category are used from the target scene. Furthermore, the baseline represents the results obtained by training solely on the target scene using 10 samples per category, without the assistance of knowledge transfer. We compare our method with six different knowledge transfer methods including Adaptor \cite{houlsby2019parameter}, Finetune \cite{lee2022exploring}, Multi-Task Learning (MTL) \cite{lee2018cross}, Domain Adaption (DA) method--UAN \cite{you2019universal} and ensemble methods--ONE \cite{zhu2018knowledge}, FFL \cite{kim2021feature}. For fair compare, we use the same backbone Masked SST \cite{scheibenreif2023masked}.  All methods are trained by the Adam optimizer with a weight decay of $5\times 10^{-3}$ and a momentum of 0.9. The initial learning rate is set to $5\times 10^{-4}$. The batch size is set to 64 for all methods.
During the transfer from IndianPine to Pavia, we set 
\(\beta = 0.1\) in Eq.~\ref{ema} and \(\tau = 2\) in Eq.~\ref{norm} under the agreement mechanism, while in the disagreement mechanism, the temperature is set to 1 in Eq.~\ref{kt_1} and \(0.05\) in Eq.~\ref{kt_2}. In addition, during the transfer from Houston to Pavia, we set 
\(\beta = 0.01\) in Eq.~\ref{ema} and \(\tau = 4\) in Eq.~\ref{norm} under the agreement mechanism, while in the disagreement mechanism, the temperature is set to 1 in Eq.~\ref{kt_1} and \(0.001\) in Eq.~\ref{kt_2}.

\subsection{Results}
\subsubsection{Compared with Knowledge Transfer Methods}
In Table \ref{Pavia}, we demonstrate that our method achieves the best results. For the majority of categories in the Pavia dataset, our method achieves the highest accuracy values. Existing methods fail to address the issues of agreement and disagreement in knowledge transfer. For example, MTL and UAN methods incorporating source scene perform similarly to or worse than the baseline transferring from Indian Pines to Pavia (I $\rightarrow$ P) and from Houston to Pavia (H $\rightarrow$ P). This suggests that merely attempting to align the source and target scenes without effectively managing gradient conflicts and dominating gradients in shared parameter is insufficient.
For disagreement, ensemble methods such as ONE and FFL increased limited performance when transferring from Indian Pines to Pavia (I $\rightarrow$ P). While these methods aim to capture different information, they fall short in significantly enhancing the model's ability to generalize to the target scene. This indicates that without disagreement restriction, the benefits of ensemble strategies remain limited.
In addition, Adaptor and Finetune methods require extra computation time (pretraining on the source scene first) to improve the performance whereas our method achieves state-of-the-art results without the need for a pretrained model. 

These results demonstrate the importance of addressing both agreement and disagreement in knowledge transfer. Our method not only overcomes the limitations of existing approaches but also provides a solution that enhances the model's ability to generalize effectively to new scenes. 

In Tables \ref{Pavia}-\ref{IndianPine}, we demonstrate that our method achieves the best results. Majority categories in Pavia, Houston and IndianPine can get the highest values in six different settings. 

Existing methods fail to address the agreement and disagreement issues. For example, for agreement, MTL and UAN methods show only slight improvements or even decreased performance when incorporating target scene, compared to the baseline when transferring from IndianPine to Pavia (I $\rightarrow$ P) and from Houston to Pavia (H $\rightarrow$ P). For disagreement,  ensemble methods including ONE and FFL increased limited performance when transferring from IndianPine to Pavia (I $\rightarrow$ P). In addition, Adaptor and Finetune methods require extra computation time (pretraining on the source scene first) to improve the performance whereas our method achieves state-of-the-art results without the need for a pretrained model. 

On the other hand, when transferring from Pavia to IndianPine (P $\rightarrow$ I) and from Houston to IndianPine (H $\rightarrow$ I), the performance of existing method decreased compared with baseline. This is mainly because most categories in IndianPine dataset are fine-grained and lack of corresponding categories in Houston and Pavia dataset. Because of this, it is difficult to increase the performance of IndianPine by utilize the information from Houston or Pavia dataset.

However, our method consider the agreement and disagreement perspective during knowledge transfer to mitigate the conflicting, dominating gradients and losing the target critical information. 



\subsubsection{Ablation Studies}
We conduct ablation studies, as shown in Table \ref{ablation}, to evaluate the importance of the agreement components (GradVac and LogitNorm) and the disagreement components (ensemble and DiR) in our method.


When Pavia (P) represent as a target scene and IndianPine (I) or Houston (H) represent as a source scene, the performance improves by 5.36\% for I $\rightarrow$ P and 6.82\% for H $\rightarrow$ P by leverage GradVac. 
Applying GradVac and LogitNorm for agreement between source scene and target scene results in a performance increase of 8.68\% for I $\rightarrow$ P and 8.21\% for H $\rightarrow$ P. Then, considering the integrate different set of target information through ensemble without disagreement restriction, the performance have limited improvement for 0.88\% for I $\rightarrow$ P and 0.25\% for H $\rightarrow$ P. Introducing the disagreement restriction, the performance can further increase 1.70\% for I $\rightarrow$ P and 0.81\% for H $\rightarrow$ P, which means it is crucial to capture a diverse aspects of target feature.

In addition, when Houston (H) work as a target scene and Pavia (P) or IndianPine (I) work as a source scene, the performance decreased by 2.11\% for P $\rightarrow$ H and 0.47\% for I $\rightarrow$ H by leverage GradVac, which means only consider the gradient conflict and neglect other issues can decrease the performance. Based on that, introducing the LogitNorm for handle the dominant gradient can improve the performance by 3.93\% for P $\rightarrow$ H and 2.34\% for I $\rightarrow$ H. It demonstrates the importance of the agreement block. In addition, the performance can increase 2.14\% for P $\rightarrow$ H and 0.80\% for I $\rightarrow$ H, utilizing the ensemble model without DiR. Leveraging the entire disagreement block can increase the performance by 3.21\% for P $\rightarrow$ H and 2.86\% for I $\rightarrow$ H.

On the other hand, when IndianPine (I) work as a target scene and Pavia (P) or Houston (H) work as a source scene, the resulting performance decreased by 2.36\% for P $\rightarrow$ I and 2.44\% for H $\rightarrow$ I by leveraging the agreement block including GradVac and LogitNorm. This is mainly because most of categories in target scene are fine-grained while source scene is coarse-grained. Furthermore, there is no clear correspondence between these fine-grained and coarse-grained categories. Introducing ensemble component can increase the performance by 3.14\% for P $\rightarrow$ I and 4.71\% for H $\rightarrow$ I. Additionally, leveraging the DiR component can further improve the performance by 1.18\% for P $\rightarrow$ I and 1.22\% for H $\rightarrow$ I.

These results demonstrate the effective by introducing the agreement and disagreement blocks. 
Moreover, when agreement and disagreement blocks are introduced together, the performance improves even further, validating the effectiveness of integrating agreement and disagreement perspectives of target information.

These results demonstrate the effectiveness of introducing both the agreement and disagreement components. Moreover, when these components are integrated together, the performance improves even further, validating the benefit of combining agreement and disagreement perspectives of target information in our method.

\begin{table*}[h]
\centering
\caption{Ablation studies on different components. $\uparrow$ indicates that higher values are better. \textbf{Bold} numbers represent the best results.}
\label{ablation}
\begin{tabular}{c|l|c}
\hline
Datasets           & GradVac \quad LogitNorm \quad ensemble \quad DiR  & OA$\uparrow$ \\ \hline
\multirow{4}{*}{I$\rightarrow$ P} &                 &    76.26              \\  
                  & \quad   \ding{51} &   81.62               \\  
                  &  \quad \ding{51}  \qquad \qquad    \ding{51}&   84.94               \\ 
                  
                  &  \quad \ding{51}  \qquad \qquad \ding{51} \qquad \qquad \ding{51}  & 85.82\\
                &  \quad \ding{51}  \qquad \qquad \ding{51} \qquad \qquad \ding{51} \qquad \qquad \ding{51}  &\textbf{87.52}\\                  \hline
\multirow{4}{*}{H$\rightarrow$ P} &                    &     73.61             \\ 
                  &  \quad   \ding{51} &   80.43               \\  
                  & \quad \ding{51}  \qquad \qquad \ding{51} &    81.82              \\ 
            &  \quad \ding{51}  \qquad \qquad \ding{51} \qquad \qquad \ding{51}  & 82.07\\
                &  \quad \ding{51}  \qquad \qquad \ding{51} \qquad \qquad \ding{51} \qquad \qquad \ding{51}  &\textbf{82.88}\\\hline

\multirow{4}{*}{P$\rightarrow$ H} &                    &     79.24             \\ 
                  &  \quad   \ding{51} &      77.13           \\  
                  & \quad \ding{51}  \qquad \qquad \ding{51} &    81.06              \\ 
            &  \quad \ding{51}  \qquad \qquad \ding{51} \qquad \qquad \ding{51}  &  83.20\\
                &  \quad \ding{51}  \qquad \qquad \ding{51} \qquad \qquad \ding{51} \qquad \qquad \ding{51}  & \textbf{84.27}\\\hline

\multirow{4}{*}{I$\rightarrow$ H} &                    &     79.24             \\ 
                  &  \quad   \ding{51} &      78.77         \\  
                  & \quad \ding{51}  \qquad \qquad \ding{51} &   81.11             \\ 
            &  \quad \ding{51}  \qquad \qquad \ding{51} \qquad \qquad \ding{51}  &  81.91  \\
                &  \quad \ding{51}  \qquad \qquad \ding{51} \qquad \qquad \ding{51} \qquad \qquad \ding{51}  &  \textbf{83.97} \\\hline
\multirow{4}{*}{P$\rightarrow$ I} &                    &     78.15             \\ 
                  &  \quad   \ding{51} &       69.20          \\  
                  & \quad \ding{51}  \qquad \qquad \ding{51} & 75.79               \\ 
            &  \quad \ding{51}  \qquad \qquad \ding{51} \qquad \qquad \ding{51}  & 78.93 \\
                &  \quad \ding{51}  \qquad \qquad \ding{51} \qquad \qquad \ding{51} \qquad \qquad \ding{51}  &\textbf{80.11} \\\hline
\multirow{4}{*}{H$\rightarrow$ I} &                    &  78.15                \\ 
                  &  \quad   \ding{51} &      67.58          \\  
                  & \quad \ding{51}  \qquad \qquad \ding{51} &  75.71              \\ 
            &  \quad \ding{51}  \qquad \qquad \ding{51} \qquad \qquad \ding{51}  & 80.42  \\
                &  \quad \ding{51}  \qquad \qquad \ding{51} \qquad \qquad \ding{51} \qquad \qquad \ding{51}  & \textbf{81.64} \\\hline
\end{tabular}
\end{table*}

\section{RELATION TO PRIOR WORK}
\label{sec:prior}
The knowledge transfer in cross-scene HSI can be broadly classified as homogeneous and heterogeneous types \cite{ning2023contrastive,ye2024adaptive,wang2024dual,ye2024cross,li2024cross, qin2024few}. For homogeneous type, the methods mainly consider the spectral shift from source to target areas in the same dataset \cite{deng2018active,ye2017dictionary,yu2021unsupervised,wang2020hyperspectral}. In heterogeneous type, the existing methods will manually match the shared categories first before aligning the distinct spectral distributions for different datasets \cite{peng2022domain, zhong2022heterogeneous, ning2023contrastive,ye2024adaptive,wang2024dual}. 
In addition, existing works primarily focus on aligning source and target domains through feature matching or distribution alignment, aiming to mitigate domain shifts.


While cross-scene hyperspectral image (HSI) transfer has seen notable progress, significant gaps persist in optimizing shared parameters and preserving diverse target features.  In cross-scene HSI scenarios, the training of shared parameters can be hindered by dominating and conflicting gradients. These gradient issues degrade performance and may lead to harmful transfer effects \cite{zhang2022survey, rosenstein2005transfer,senushkin2023independent}. Approaches such as GradNorm \cite{wang2020gradient} and PCGrad \cite{yu2020gradient} have been proposed in multi-task learning to align gradients and prevent conflicts when updating shared parameters. However, these techniques are not specifically designed to address the unique challenges of cross-scene HSI transfer. In addition, current approaches tend to rely on a shared subset of target features for predictive outcomes. This reliance can result in the potential loss of critical target information, thereby preventing the model from fully capturing the rich and diverse patterns present in the target scene \cite{pagliardini2023agree}. Outside HSI, knowledge distillation has shown promise in leveraging different teacher models to capture a broader range of target perspectives \cite{hinton2015distilling,allen2020towards,chen2020online}. Despite these advancements, there remains an unmet need for a unified framework that simultaneously addresses both the optimization of shared parameters and the preservation of diverse target features within the challenging context of cross-scene HSI transfer.

\section{Discussion}


In this paper, we introduce the agreement-disagreement guided knowledge transfer (ADGKT) method that integrates both agreement and disagreement mechanisms to enhance knowledge transfer between cross-scene HSI. The agreement mechanisms, consisting of GradVac and LogitNorm, effectively address gradient conflicts and dominating gradients by aligning gradient directions and controlling the magnitude of logits, balancing shared parameter optimization. The disagreement mechanisms, incorporating a DiR and an ensemble approach, capturing a diverse and independent target features. This mechanisms mitigates the risk of losing critical target information that might be overlooked when focusing solely on agreement between source and target scenes. 
Although our method demonstrates state-of-the-art performance, we acknowledge several considerations for future exploration. First, the method involves a few hyperparameters that may require systematic tuning to optimize performance in different datasets. Second, incorporating both agreement and
disagreement mechanisms, along with an ensemble strategy, can lead to moderately increased training complexity compared to simpler baselines.

\section{Conclusions}

In this paper, we propose the agreement-disagreement guided knowledge transfer (ADGKT) method, which integrates both agreement and disagreement mechanisms to enhance knowledge transfer across cross-scene HSI. The agreement mechanism, consisting of GradVac and LogitNorm, effectively address gradient conflicts and dominating gradients by aligning gradient directions and controlling the magnitude of logits, balancing shared parameter optimization. The disagreement mechanism, incorporating a DiR and an ensemble approach, capturing a diverse and independent target features. This mechanism mitigates the risk of losing critical target information that might be overlooked when focusing solely on agreement between source and target scenes. Our extensive experiments demonstrate that ADGKT achieves state-of-the-art performance, outperforming existing methods that fail to adequately address both agreement and disagreement issues.

\bibliography{sn-bibliography}

@article{ghamisi2017advances,
  title={Advances in hyperspectral image and signal processing: A comprehensive overview of the state of the art},
  author={Ghamisi, Pedram and Yokoya, Naoto and Li, Jun and Liao, Wenzhi and Liu, Sicong and Plaza, Javier and Rasti, Behnood and Plaza, Antonio},
  journal={IEEE Geoscience and Remote Sensing Magazine},
  volume={5},
  number={4},
  pages={37--78},
  year={2017},
  publisher={IEEE}
}

@inproceedings{Landgrebe2002HyperspectralID,
  title={Hyperspectral Image Data Analysis as a High Dimensional Signal Processing Problem},
  author={David A. Landgrebe},
  year={2002},
  url={https://api.semanticscholar.org/CorpusID:56804707}
}

@article{ye2017dictionary,
  title={Dictionary learning-based feature-level domain adaptation for cross-scene hyperspectral image classification},
  author={Ye, Minchao and Qian, Yuntao and Zhou, Jun and Tang, Yuan Yan},
  journal={IEEE Transactions on Geoscience and Remote Sensing},
  volume={55},
  number={3},
  pages={1544--1562},
  year={2017},
  publisher={IEEE}
}

@article{lee2022exploring,
  title={Exploring cross-domain pretrained model for hyperspectral image classification},
  author={Lee, Hyungtae and Eum, Sungmin and Kwon, Heesung},
  journal={IEEE Transactions on Geoscience and Remote Sensing},
  volume={60},
  pages={1--12},
  year={2022},
  publisher={IEEE}
}

@article{deng2018active,
  title={Active multi-kernel domain adaptation for hyperspectral image classification},
  author={Deng, Cheng and Liu, Xianglong and Li, Chao and Tao, Dacheng},
  journal={Pattern Recognition},
  volume={77},
  pages={306--315},
  year={2018},
  publisher={Elsevier}
}

@article{yu2021unsupervised,
  title={Unsupervised domain adaptation with dense-based compaction for hyperspectral imagery},
  author={Yu, Chunyan and Liu, Caiyu and Yu, Haoyang and Song, Meiping and Chang, Chein-I},
  journal={IEEE Journal of Selected Topics in Applied Earth Observations and Remote Sensing},
  volume={14},
  pages={12287--12299},
  year={2021},
  publisher={IEEE}
}

@article{peng2022domain,
  title={Domain adaptation in remote sensing image classification: A survey},
  author={Peng, Jiangtao and Huang, Yi and Sun, Weiwei and Chen, Na and Ning, Yujie and Du, Qian},
  journal={IEEE Journal of Selected Topics in Applied Earth Observations and Remote Sensing},
  volume={15},
  pages={9842--9859},
  year={2022},
  publisher={IEEE}
}

@inproceedings{you2019universal,
  title={Universal domain adaptation},
  author={You, Kaichao and Long, Mingsheng and Cao, Zhangjie and Wang, Jianmin and Jordan, Michael I},
  booktitle={Proceedings of the IEEE/CVF conference on computer vision and pattern recognition},
  pages={2720--2729},
  year={2019}
}

@inproceedings{zhen2022versatile,
  title={On the versatile uses of partial distance correlation in deep learning},
  author={Zhen, Xingjian and Meng, Zihang and Chakraborty, Rudrasis and Singh, Vikas},
  booktitle={European Conference on Computer Vision},
  pages={327--346},
  year={2022},
  organization={Springer}
}

@article{li2022shadow,
  title={Shadow knowledge distillation: Bridging offline and online knowledge transfer},
  author={Li, Lujun and Jin, Zhe},
  journal={Advances in Neural Information Processing Systems},
  volume={35},
  pages={635--649},
  year={2022}
}

@inproceedings{houlsby2019parameter,
  title={Parameter-efficient transfer learning for NLP},
  author={Houlsby, Neil and Giurgiu, Andrei and Jastrzebski, Stanislaw and Morrone, Bruna and De Laroussilhe, Quentin and Gesmundo, Andrea and Attariyan, Mona and Gelly, Sylvain},
  booktitle={International Conference on Machine Learning},
  pages={2790--2799},
  year={2019},
  organization={PMLR}
}

@article{hinton2015distilling,
  title={Distilling the knowledge in a neural network},
  author={Hinton, Geoffrey and Vinyals, Oriol and Dean, Jeff},
  journal={arXiv preprint arXiv:1503.02531},
  year={2015}
}

@article{zhu2018knowledge,
  title={Knowledge distillation by on-the-fly native ensemble},
  author={Zhu, Xiatian and Gong, Shaogang and others},
  journal={Advances in neural information processing systems},
  volume={31},
  year={2018}
}

@inproceedings{kim2021feature,
  title={Feature fusion for online mutual knowledge distillation},
  author={Kim, Jangho and Hyun, Minsung and Chung, Inseop and Kwak, Nojun},
  booktitle={2020 25th International Conference on Pattern Recognition (ICPR)},
  pages={4619--4625},
  year={2021},
  organization={IEEE}
}

@inproceedings{lee2018cross,
  title={Cross-domain CNN for hyperspectral image classification},
  author={Lee, Hyungtae and Eum, Sungmin and Kwon, Heesung},
  booktitle={IGARSS 2018-2018 IEEE International Geoscience and Remote Sensing Symposium},
  pages={3627--3630},
  year={2018},
  organization={IEEE}
}

@inproceedings{scheibenreif2023masked,
  title={Masked vision transformers for hyperspectral image classification},
  author={Scheibenreif, Linus and Mommert, Michael and Borth, Damian},
  booktitle={Proceedings of the IEEE/CVF Conference on Computer Vision and Pattern Recognition},
  pages={2165--2175},
  year={2023}
}

@article{zhang2022survey,
  title={A survey on negative transfer},
  author={Zhang, Wen and Deng, Lingfei and Zhang, Lei and Wu, Dongrui},
  journal={IEEE/CAA Journal of Automatica Sinica},
  volume={10},
  number={2},
  pages={305--329},
  year={2022},
  publisher={IEEE}
}

@inproceedings{rosenstein2005transfer,
  title={To transfer or not to transfer},
  author={Rosenstein, Michael T and Marx, Zvika and Kaelbling, Leslie Pack and Dietterich, Thomas G},
  booktitle={NIPS 2005 workshop on transfer learning},
  volume={898},
  number={3},
  year={2005}
}

@inproceedings{
pagliardini2023agree,
title={Agree to Disagree: Diversity through Disagreement for Better Transferability},
author={Matteo Pagliardini and Martin Jaggi and Fran{\c{c}}ois Fleuret and Sai Praneeth Karimireddy},
booktitle={The Eleventh International Conference on Learning Representations },
year={2023},
url={https://openreview.net/forum?id=K7CbYQbyYhY}
}

@inproceedings{senushkin2023independent,
  title={Independent component alignment for multi-task learning},
  author={Senushkin, Dmitry and Patakin, Nikolay and Kuznetsov, Arseny and Konushin, Anton},
  booktitle={Proceedings of the IEEE/CVF Conference on Computer Vision and Pattern Recognition},
  pages={20083--20093},
  year={2023}
}

@article{wang2020hyperspectral,
  title={Hyperspectral image classification based on domain adaptation broad learning},
  author={Wang, Haoyu and Wang, Xuesong and Chen, CL Philip and Cheng, Yuhu},
  journal={IEEE Journal of Selected Topics in Applied Earth Observations and Remote Sensing},
  volume={13},
  pages={3006--3018},
  year={2020},
  publisher={IEEE}
}

@article{ye2024adaptive,
  title={Adaptive graph modeling with self-training for heterogeneous cross-scene hyperspectral image classification},
  author={Ye, Minchao and Chen, Junbin and Xiong, Fengchao and Qian, Yuntao},
  journal={IEEE Transactions on Geoscience and Remote Sensing},
  year={2024},
  publisher={IEEE}
}

@article{zhong2022heterogeneous,
  title={Heterogeneous spectral-spatial feature transfer with structure preserved distribution alignment for hyperspectral image classification},
  author={Zhong, Chongxiao and Zhang, Junping and Guo, Qingle and Zhang, Ye},
  journal={IEEE Journal of Selected Topics in Applied Earth Observations and Remote Sensing},
  volume={15},
  pages={5545--5558},
  year={2022},
  publisher={IEEE}
}

@article{allen2020towards,
  title={Towards understanding ensemble, knowledge distillation and self-distillation in deep learning},
  author={Allen-Zhu, Zeyuan and Li, Yuanzhi},
  journal={arXiv preprint arXiv:2012.09816},
  year={2020}
}

@inproceedings{chen2020online,
  title={Online knowledge distillation with diverse peers},
  author={Chen, Defang and Mei, Jian-Ping and Wang, Can and Feng, Yan and Chen, Chun},
  booktitle={Proceedings of the AAAI conference on artificial intelligence},
  volume={34},
  number={04},
  pages={3430--3437},
  year={2020}
}

@article{ning2023contrastive,
  title={Contrastive learning based on category matching for domain adaptation in hyperspectral image classification},
  author={Ning, Yujie and Peng, Jiangtao and Liu, Quanyong and Huang, Yi and Sun, Weiwei and Du, Qian},
  journal={IEEE Transactions on Geoscience and Remote Sensing},
  year={2023},
  publisher={IEEE}
}

@article{li2024cross,
  title={Cross-Domain Few-shot Hyperspectral Image Classification with Cross-Modal Alignment and Supervised Contrastive Learning},
  author={Li, Zhaokui and Zhang, Chenyang and Wang, Yan and Li, Wei and Du, Qian and Fang, Zhuoqun and Chen, Yushi},
  journal={IEEE Transactions on Geoscience and Remote Sensing},
  year={2024},
  publisher={IEEE}
}

@article{qin2024few,
  title={Few-Shot Learning with Prototype Rectification for Cross-Domain Hyperspectral Image Classification},
  author={Qin, Anyong and Yuan, Chaoqi and Li, Qiang and Luo, Xiaoliu and Yang, Feng and Song, Tiecheng and Gao, Chenqiang},
  journal={IEEE Transactions on Geoscience and Remote Sensing},
  year={2024},
  publisher={IEEE}
}

@article{wang2024dual,
  title={Dual-Branch Domain Adaptation Few-Shot Learning for Hyperspectral Image Classification},
  author={Wang, Zhuowei and Zhao, Shihui and Zhao, Genping and Song, Xiaoyu},
  journal={IEEE Transactions on Geoscience and Remote Sensing},
  year={2024},
  publisher={IEEE}
}

@article{ye2024cross,
  title={Cross-domain few-shot learning based on graph convolution contrast for hyperspectral image classification},
  author={Ye, Zhen and Wang, Jie and Sun, Tao and Zhang, Jinxin and Li, Wei},
  journal={IEEE Transactions on Geoscience and Remote Sensing},
  year={2024},
  publisher={IEEE}
}

@article{day2017survey,
  title={A survey on heterogeneous transfer learning},
  author={Day, Oscar and Khoshgoftaar, Taghi M},
  journal={Journal of Big Data},
  volume={4},
  pages={1--42},
  year={2017},
  publisher={Springer}
}

@inproceedings{wei2022mitigating,
  title={Mitigating neural network overconfidence with logit normalization},
  author={Wei, Hongxin and Xie, Renchunzi and Cheng, Hao and Feng, Lei and An, Bo and Li, Yixuan},
  booktitle={International conference on machine learning},
  pages={23631--23644},
  year={2022},
  organization={PMLR}
}

@article{wang2020gradient,
  title={Gradient vaccine: Investigating and improving multi-task optimization in massively multilingual models},
  author={Wang, Zirui and Tsvetkov, Yulia and Firat, Orhan and Cao, Yuan},
  journal={arXiv preprint arXiv:2010.05874},
  year={2020}
}

@article{yu2020gradient,
  title={Gradient surgery for multi-task learning},
  author={Yu, Tianhe and Kumar, Saurabh and Gupta, Abhishek and Levine, Sergey and Hausman, Karol and Finn, Chelsea},
  journal={Advances in Neural Information Processing Systems},
  volume={33},
  pages={5824--5836},
  year={2020}
}

@article{flarence2025hyperspectral,
  title={Hyperspectral image classification using modified mutual nearest neighbour clustering},
  author={Flarence, R Aruna and Rupa, B and Negi, Atul},
  journal={Multimedia Tools and Applications},
  pages={1--33},
  year={2025},
  publisher={Springer}
}

@article{dabas2025construction,
  title={Construction of hyperspectral images from RGB images via CNN},
  author={Dabas, Vibhuti and Jaiswal, Garima and Agarwal, Mohit and Rani, Ritu and Sharma, Arun},
  journal={Multimedia Tools and Applications},
  volume={84},
  number={11},
  pages={8725--8744},
  year={2025},
  publisher={Springer}
}

\end{document}